\pdfoutput=1
\documentclass[a4paper,12pt]{article}
\usepackage{amsfonts}
\usepackage{amssymb}
\usepackage{amsmath}
\usepackage{graphicx}
\usepackage{authblk}
\usepackage{bbold}

\usepackage[psnames]{xcolor}

\usepackage{subfigure}

\newcommand{\ket}[1]{| #1 \rangle}
\newcommand{\bra}[1]{\langle #1 |}
\newcommand{\ketbra}[2]{| #1 \rangle \langle #2 |}

\newcommand{\tr}{\mathrm{tr}}
\newcommand{\dd}{\mathrm{d}}
\newcommand{\ii}{\mathrm{i}}
\newcommand{\1}{\mathbb{1}}

\title{Decoherence effects in the quantum qubit flip game using Markovian 
approximation}
\date{25/06/2013}
\author[$*$]{Piotr Gawron}
\author[$*$,$\dagger$]{Dariusz Kurzyk}
\author[$*$]{{\L}ukasz Pawela}
\affil[$*$]{Institute of Theoretical and Applied Informatics, Polish Academy
of Sciences, Ba{\l}tycka 5, 44-100 Gliwice, Poland}
\affil[$\dagger$]{Institute of Mathematics, Silesian University of Technology, 
Kaszubska 23, Gliwice 44-100, Poland}

\begin{document}
\maketitle

\begin{abstract}
We are considering a quantum version of the penny flip game, whose
implementation is influenced by the environment that causes decoherence of the
system. In order to model the decoherence we assume Markovian approximation of
open quantum system dynamics. We focus our attention on the phase damping,
amplitude damping and amplitude raising channels. Our results show that the
Pauli strategy is no longer a Nash equilibrium under decoherence. We attempt to
optimize the players' control pulses in the aforementioned setup to allow them
to achieve higher probability of winning the game compared to the Pauli
strategy.

\end{abstract}

\section{Introduction}
Quantum information experiments can be described as a sequence of three
operations: state preparation, evolution and measurement
\cite{heinosaari2011mathematical}. In most cases one cannot assume that
experiments are conducted perfectly, therefore imperfections have to be taken
into account while modelling them. In this work we are interested in how the
knowledge about imperfect evolution of a quantum system can be exploited  by
players engaged in a quantum game. We assume that one of the players  possesses
the knowledge about imperfections in the system, while the other is  ignorant
of their existence. We ask a question of how much the player's knowledge  about
those imperfections can be exploited by him/her for their advantage.

We consider implementation of the quantum version of the penny flip game, which
is influenced by the environment that causes decoherence of
the system. In order to model the decoherence we assume Markovian approximation
of open quantum system dynamics.

The paper is organised as follows: in the two following subsections we discuss
related work and present our motivation to undertake this task. In
Section~\ref{sec:game} we recall the penny flip game and its quantum version,
in Section~\ref{sec:master-equ} we present the noise model, in
Section~\ref{sec:strategies} we discuss the strategies applied in  the presence
of noise and finally in Section~\ref{sec:conclusions} we conclude  the obtained
results.

\subsection{Related work}
Imperfect realizations of quantum games have been discussed in literature since
the beginning of the century. Ref.~\cite{johnson2001playing} discusses a
three-player quantum game played with a corrupted source of entangled qubits.
The author implicitly assumes that the initial state of the game had passed
through a bit-flip noisy channel before the game began. The corruption of
quantum states in schemes implementing quantum games was studied by various
authors \textit{i.e.} in \cite{chen2002noisy} the authors perform an analysis
of the two-player prisoners dilemma game, in \cite{flitney2005multiplayer} the
multiplayer quantum minority game with decoherence is studied, in
\cite{gawron2008noise, pawela2013enhancing} the authors analyse the influence
of the local noisy channels on quantum Magic Squares games, while the quantum
Monty Hall problem under decoherence is studied first in \cite{gawron2010noisy}
and subsequently in \cite{khan2010quantum}. In \cite{miszczak2012qubit} the
authors study the influence of the interaction of qubits forming a spin chain
on the qubit flip game. An analysis of trembling hand perfect equilibria in
quantum games was done in \cite{pakula2008analysis}. Prisoners' dilemma in the
presence of collective dephasing modelled by using the Markovian approximation
of open quantum systems dynamics is studied in \cite{nawaz2012prisoners}.
Unfortunately the model applied in this work assumes that decoherence acts only
after the  initial state has been prepared and ceases to act before unitary
strategies are  applied.

\subsection{Motivation}
In the quantum game theoretic literature decoherence is typically applied to a
quantum game in the following way:
\begin{enumerate}
\item the entangled state is prepared,
\item it is transferred through a local noisy channel,
\item players' strategies are applied,
\item \label{item:decoherence2} the resulting state is transferred once again
through a local noisy channel,
\item \label{item:disentanglement} the state is disentangled,
\item quantum local measurements are performed and the outcomes of the games
are calculated.
\end{enumerate}
In some cases, where it is appropriate, steps \ref{item:decoherence2} and
\ref{item:disentanglement} are omitted. The problem with the above procedure is
that it separates unitary evolution from the decoherent evolution. In
\cite{miszczak2012qubit} it was proposed to observe the behaviour of the
quantum version of the penny flip game under more physically realistic
assumptions where decoherence due to coupling with the environment and unitary
evolution happen simultaneously.

\section{Game as a quantum experiment}\label{sec:game}

In this work our goal is to follow the work done in \cite{miszczak2012qubit}
and to discuss the quantum penny flip game as a physical experiment consisting
in preparation, evolution and measurement of the system. For the purpose of
this paper we assume that preparation and measurement, contrary to noisy
evolution of the system are perfect. We investigate the influence of the noise
on the players' odds and how the noisiness of the system can be exploited by
them. The noise model we use is described by the Lindblad master
equation  and the dynamics of the system is expressed in the language of
quantum systems control.

\subsection{Penny flip game}
In order to provide classical background for our problem, let us consider a
classical two-player game, consisting in flipping over a coin by the players
in three consecutive rounds. As usual, the players are called Alice and Bob. In
each round Alice and Bob performs one of two operations on the coin:
flips it over or retains it unchanged.

At the beginning of the game, the coin is turned heads up. During the course of
the game the coin is hidden and the players do not know the opponents actions.
If after the last round the coin is tails up, then Alice wins, otherwise the
winner is Bob.

The game consists of three rounds: Alice performs her action in the first and
the third round, while Bob performs his in the second round of the game.
Therefore the set of allowed strategies consists of eight sequences $(N,N,N),
(N,N,F),$ $ \ldots, (F,F,F)$, where $N$ corresponds to the \emph{non-flipping
strategy} and $F$ to the \emph{flipping strategy}. Bob's pay-off table for this
game is presented in Table  \ref{tab:classic-payoff-table}. Looking at the
pay-off tables, it can be seen that utility function of players in the game is
balanced, thus the penny flip game is  a zero-sum game.

\begin{table}[h]
\centering
\begin{tabular}{ |c|c|c|c|c| }
 \hline            
  & $NN$ & $FN$ & $NF$ & $FF$ \\
 \hline 
 $N$ & $1$ & $-1$ & $-1$ & $1$ \\
 \hline 
 $F$ & $-1$ & $1$ & $1$ & $-1$ \\
 \hline 
\end{tabular}
\caption{Bob's pay-off table for the penny flip game.}
\label{tab:classic-payoff-table}
\end{table}

A detailed analysis of this game and its asymmetrical quantization can be found
in \cite{piotrowski2003invitation}. In this work it was shown that there is no
winning strategy for any player in the penny flip game. It was also shown, that
if Alice was allowed to extend her set of strategies to quantum strategies she
could always win. In \cite{miszczak2012qubit} it was shown that when both
players have access to quantum strategies the game becomes fair and it has the
Nash equilibrium.

\subsection{Qubit flip game}
Following the work done in the aforementioned paper \cite{miszczak2012qubit} we
consider a quantum version of the penny flip game. In this case, we treat a
qubit  as a quantum coin. As in the classical case the game is divided into
three rounds. Starting with Alice, in each round, one player performs a unitary
operation on the quantum coin. The rules of the game are  constrained by its
physical implementation. We assume that in each round each of the players can
choose three control parameters  $\alpha_1,\alpha_2,\alpha_3$ in order to
realise his/her strategy. The  resulting unitary gate is given by the equation:
\begin{equation}
U(\alpha_1,\alpha_2,\alpha_3)=e^{-\ii\alpha_3\sigma_z \Delta t}
e^{-\ii\alpha_2\sigma_y \Delta t} e^{-\ii\alpha_1\sigma_z \Delta t},
\end{equation} 
where $\Delta t$ is an arbitrarily chosen constant time interval.

Therefore, the system defined above forms a single qubit system driven by
time-dependent Hamiltonian $H(t)$, which is a piecewise constant and can be
expressed in the following form
\begin{equation}\label{equ:hamiltonian}
H(t)=
\begin{cases}
\alpha_1^{A_1}\sigma_z & \text{ for } 0\leq t < \Delta t,\\
\alpha_2^{A_1}\sigma_y & \text{ for } \Delta t\leq t < 2\Delta t,\\
\alpha_3^{A_1}\sigma_z & \text{ for } 2\Delta t\leq t < 3\Delta t,\\
\alpha_1^{B}\sigma_z & \text{ for } 3\Delta t\leq t < 4\Delta t,\\
\alpha_2^{B}\sigma_y & \text{ for } 4\Delta t\leq t < 5\Delta t,\\
\alpha_3^{B}\sigma_z & \text{ for } 5\Delta t\leq t < 6\Delta t,\\
\alpha_1^{A_2}\sigma_z & \text{ for } 6\Delta t\leq t < 7\Delta t,\\
\alpha_2^{A_2}\sigma_y & \text{ for } 7\Delta t\leq t < 8\Delta t,\\
\alpha_3^{A_2}\sigma_z & \text{ for } 8\Delta t\leq t \leq 9\Delta t.
\end{cases}
\end{equation}
Control parameters in the Hamiltonian $H(t)$ will be referred to vector
$\mathrm{\alpha}=(\alpha_1^{A_1}, \alpha_1^{A_1}, \alpha_2^{A_1},
\alpha_2^{A_1}, \alpha_1^{B}, \alpha_2^{B}, \alpha_3^{B}, \alpha_1^{A_2},
\alpha_2^{A_2}, \alpha_3^{A_2})$, where $\alpha_i^{A_j}$ are determined by
Alice and $\alpha_i^{B}$ are selected by Bob.

Suppose that players are allowed to play the game by manipulating the control
parameters in the Hamiltonian $H(t)$ representing the coherent part of the
dynamics, but they are not aware of the action of the environment on the
system. Hence the time evolution of the system is non-unitary and is described
by a master equation, which can be written generally in the  \emph{Lindblad}
form as
\begin{equation}
\frac{\dd\rho}{\dd t}=-\ii[H(t),\rho] +
\sum_j \gamma_j(L_j\rho L_j^\dagger -
\frac{1}{2}\{L_j^\dagger L_j,\rho\}),
\label{equ:master-equation}
\end{equation} 
where $H(t)$ is the system Hamiltonian, $L_j$ are the \emph{Lindblad}
operators, representing the environment influence on the system
\cite{nielsen2000quantum} and $\rho$ is the state of the system.

For the purpose of this paper we chose three classes of decoherence:
\emph{amplitude damping}, \emph{amplitude raising} and \emph{phase damping}
which correspond to noisy operators $\sigma_{-}=\ketbra{0}{1}$,
$\sigma_{+}=\ketbra{1}{0}$ and $\sigma_z$, respectively.

Let us suppose that initially the quantum coin is in the state $\ketbra{0}{0}$.
Next, in each round, Alice and Bob perform their sequences of controls on the
qubit, where each control pulse is applied  according to equation
(\ref{equ:master-equation}). After applying all of the nine pulses, we
measure the expected value of the $\sigma_z$ operator. If
$\tr(\sigma_z\rho(T))=-1$ Alice wins, if $\tr(\sigma_z\rho(T))=1$ Bob wins.
Here, $\rho(T)$ denotes the state of the system at time $T=9\Delta t$.

Alternatively we can say that the final step of the procedures consists in
performing orthogonal measurement$\{O_{\rm heads}\to\ketbra{0}{0},O_{\rm
tails}\to\ketbra{1}{1}\}$ on state $\rho(T)$. The probability of measuring
$O_{\rm heads}$ and $O_{\rm tails}$ determines pay-off functions for Alice and
Bob, respectively. These probabilities can be obtained from relations
$p(\mathrm{heads})=\bra{1}\rho(T)\ket{1}$ and
$p(\mathrm{tails})=\bra{0}\rho(T)\ket{0}$.

\subsection{Nash equilibrium}
In this game, pure strategies cannot be in Nash equilibrium. Hence, the players
choose mixed strategies, which are better than the pure ones. We assume that
Alice and Bob use the \emph{Pauli strategy}, which is mixed and gives Nash
equilibrium  \cite{miszczak2012qubit}, therefore this strategy is a reasonable
choice for the players.  According to the Pauli strategy, each player chooses
one of the four unitary operations  $\{\1, \ii\sigma_{x}, \ii\sigma_{y},
\ii\sigma_{z}\}$ with equal probability.  Thus, to obtain the Pauli strategy,
each player chooses a sequence of control  parameters $(\xi_1, \xi_1, \xi_2)$
listed  in Tab.~\ref{tab:control-parameter}.  It means that in each round, one
player performs a unitary operation  chosen randomly with a uniform probability
distribution from the set $\{ \1,  \ii\sigma_x, \ii\sigma_y, \ii\sigma_z \}$.

\begin{table}[h]
\centering
\begin{tabular}{ |c|c|c|c| }
\hline
& $\xi_1$ & $\xi_2$ & $\xi_3$ \\
\hline
$\1$ & $0$ & $0$ & $0$ \\
\hline
$\ii\sigma_x$ & $\frac{\pi}{4}$ & $-\frac{\pi}{2}$ & $-\frac{\pi}{4}$ \\
\hline
$\ii\sigma_y$ & $0$ & $-\frac{\pi}{2}$ & $0$ \\
\hline
$\ii\sigma_z$ & $-\frac{\pi}{4}$ & $0$ & $-\frac{\pi}{4}$ \\
\hline
\end{tabular}
\caption{Control parameters for realising the Pauli strategy. The left column
indicates the resulting gate.}
\label{tab:control-parameter}
\end{table}

\section{Influence of decoherence on the game}\label{sec:master-equ}
In this section, we perform an analytical investigation which shows the
influence of decoherence on the game result. In accordance with the Lindblad
master equation, the environment influence on the system is represented by
Lindblad operators $L_j$, while the rate of decoherence is described by
parameters $\gamma_j$. To simplify the discussion, we consider Hamiltonians
$H(t)$ represented by diagonal matrices, i.e. in the following form
$H(t)=\beta_1\ketbra{0}{0}+\beta_2\ketbra{1}{1}$.

\subsection{Amplitude damping and amplitude raising}
First we consider the  amplitude damping decoherence, which corresponds to the
Lindblad operator $\sigma_{-}$.
Thus the master equation (\ref{equ:master-equation}) is expressed as
\begin{equation} \frac{\dd\rho}{\dd t}=-\ii[H(t),\rho] +
\gamma(\sigma_{-}\rho\sigma_{+}-\frac{1}{2}\sigma_{+}\sigma_{-}\rho-\frac{1}{2}\
\rho\sigma_{+}\sigma_{-}), \end{equation}
where $\sigma_{+}=\sigma_{-}^\dagger=\ketbra{1}{0}$. The equation can be
rewritten in the following form
\begin{equation}
\frac{\dd\rho}{\dd t}=A\rho+\rho A^\dagger+ \gamma\sigma_{-}\rho\sigma_{+},
\end{equation}
where $A=-\ii H(t)-\frac{1}{2}\gamma\sigma_{+}\sigma_{-}$. In solving this
equation it is helpful to make a change of variables
$\rho(t)=e^{At}\hat{\rho}(t)e^{A^\dagger t}$. Hence, we  obtain
\begin{equation}
\frac{\dd\hat{\rho}}{\dd t}=\gamma B(t)\hat{\rho}(t) B^{\dagger}(t),
\end{equation}
where
$B(t)=e^{-At}\sigma_{-}e^{At}=e^{-\ii(\beta_2-\beta_1)t-\frac{\gamma}{2}t}\sigma_{-}$. It follows that
\begin{equation}\label{eqAfterTransform}
\frac{\dd\hat{\rho}}{\dd t}=\gamma e^{-\gamma t} \sigma_{-}\hat{\rho}(t)
\sigma_{+}.
\end{equation}
Due to the fact that $\sigma_{-}\sigma_{-}=\sigma_{+}\sigma_{+}=0$ and
$\sigma_{-}\frac{\dd\hat{\rho}}{\dd t}\sigma_{+}=0$ it is possible to write
$\hat{\rho}(t)$ as
\begin{equation}
\hat{\rho}(t)=\hat{\rho}(0) -e^{-\gamma t}\sigma_{-}\hat{\rho}(0)\sigma_{+}.
\end{equation}
Coming back to the original variables we get the expression
\begin{equation}
\rho(t)=e^{At}\rho(0)e^{A^\dagger
t}-e^{-\gamma t}\sigma_{-}\rho(0)\sigma_{+}.
\end{equation}

\begin{figure}[p!]
   \centering
   \subfigure[Control parameters\newline $\alpha=(-\frac{\pi}{4},-\frac{\pi}{2},\frac{\pi}{4},0,-\frac{\pi}{2},0,-\frac{\pi}{4},-\frac{\pi}{2},\frac{\pi}{4})$.
    ]{%
     \centering
     \includegraphics{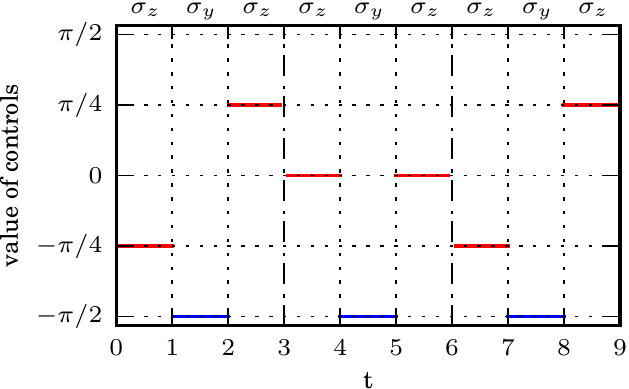}
     \label{ad_:a}
   }%
   \subfigure[Control parameters\newline $\alpha=(0,-\frac{\pi}{2},0,-\frac{\pi}{4},-\frac{\pi}{2},\frac{\pi}{4},-\frac{\pi}{4},0,-\frac{\pi}{4})$.
        ]{%
     \centering
     \includegraphics{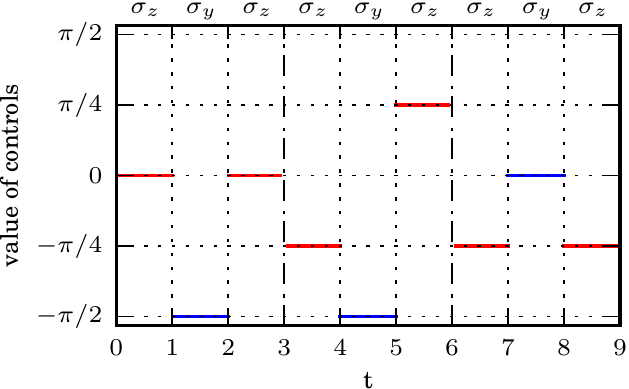}
     \label{ad_:b}
   }%
   \\
    \subfigure[Mean values of $\sigma_x,\sigma_y$ and $\sigma_z$.
     ]{%
      \centering
      \includegraphics{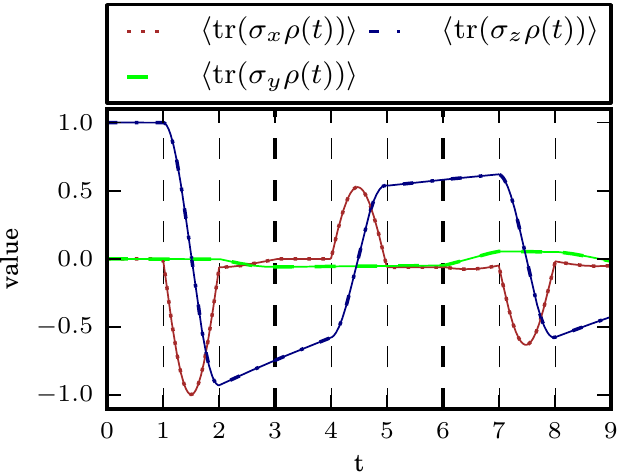}
      \label{ad_:c}
    }%
    \subfigure[Mean values of $\sigma_x,\sigma_y$ and $\sigma_z$. 
         ]{%
      \centering
      \includegraphics{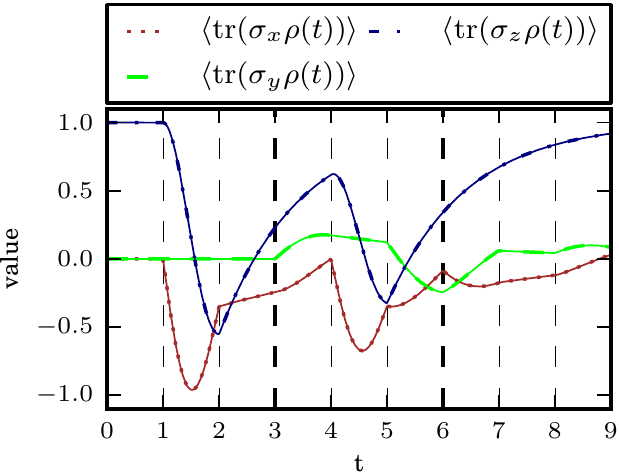}
      \label{ad_:d}
    }%
    \\
    \subfigure[Time evolution of a quantum coin.
         ]{%
    	\centering
      \includegraphics{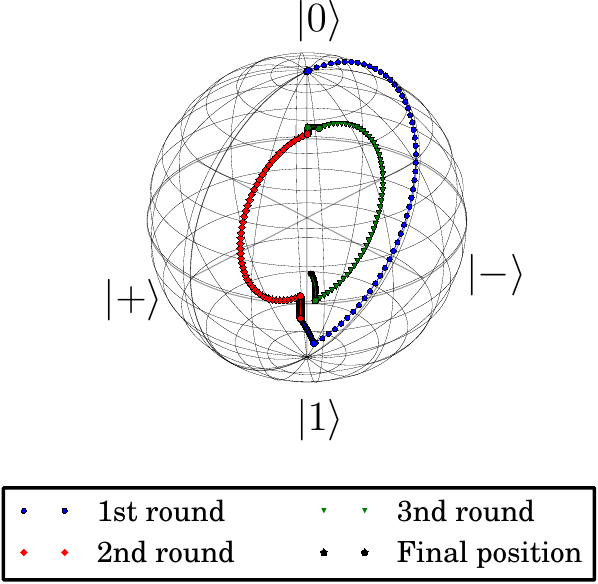}
      \label{ad_:e}
    }%
    \subfigure[Time evolution of a quantum coin.
         ]{%
      \centering
      \includegraphics{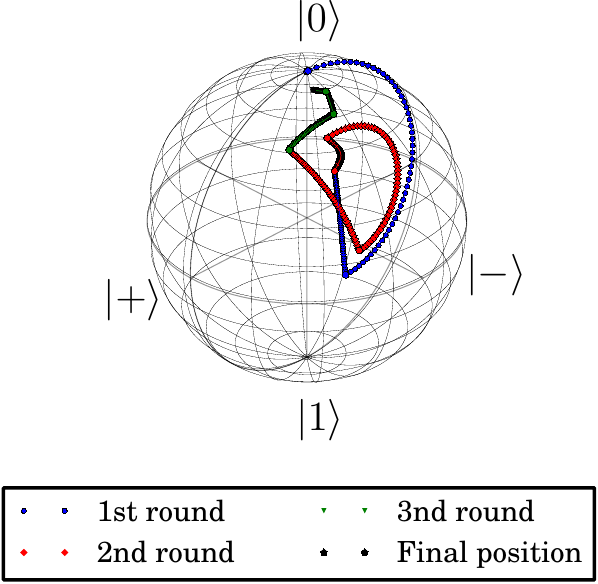}
      \label{ad_:f}
    }%

  \caption{Example of the time evolution of a quantum system with the amplitude
damping decoherence for a sequence of control parameters $\alpha$ and  fixed
$\gamma=0.1$ (left side), $\gamma=0.7$ (right side).} %
\label{fig:evolution_ad}
\end{figure}

In order to study the asymptotic effects of decoherence on the results of the
game we consider the following limit
\begin{equation}
\lim_{\gamma\to\infty} e^{At}\rho(0)e^{A^\dagger
t}-e^{-\gamma t}\sigma_{+}\rho(0)\sigma_{-} = \ketbra{0}{0}\rho(0)\ketbra{0}{0}.
\end{equation}
Let $\rho(0)=\ketbra{0}{0}$, thus the above limit is equal to $\ketbra{0}{0}$.
This result shows that for high values of $\gamma$, chances of winning the game
by Bob increase to 1 as $\gamma$ increases. Figure \ref{fig:evolution_ad}
shows an example of the evolution of a quantum  system with amplitude damping
decoherence.

The noisy operator $\sigma_{+}$ is related to amplitude raising decoherence,
and the solution of the master equation has the following form
\begin{equation}
\rho(t)=e^{At}\rho(0)e^{A^\dagger 
t}-e^{-\gamma t}\sigma_{+}\rho(0)\sigma_{-},
\end{equation}
where $A=-\ii H(t) -\frac{1}{2}\gamma\sigma_{-}\sigma_{+}$. It is easy to check
that as $\gamma \rightarrow \infty$ the state $\ketbra{1}{1}$ is the solution
of the above equation, in which case Alice wins.

\subsection{Phase damping}
\begin{figure}[p!]
   \centering
    \subfigure[Control parameters\newline$\alpha=(-\frac{\pi}{4},-\frac{\pi}{2},\frac{\pi}{4},-\frac{\pi}{4},-\frac{\pi}{2},\frac{\pi}{4},-\frac{\pi}{4},-\frac{\pi}{2},\frac{\pi}{4})$.
     ]{%
      \centering
      \includegraphics{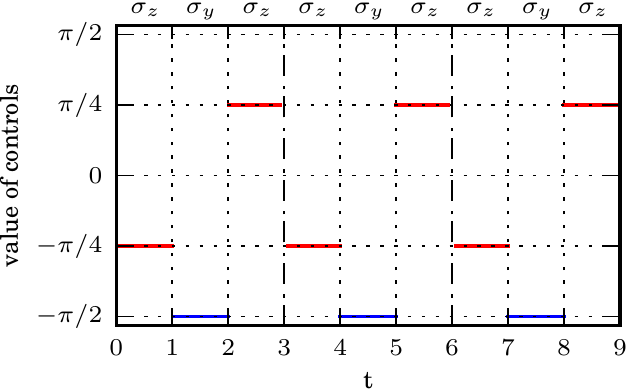}
      \label{pd_:a}
    }%
    \subfigure[Control parameters\newline$\alpha=(0,-\frac{\pi}{2},0,0,-\frac{\pi}{2},0,0,-\frac{\pi}{2},0)$.
         ]{%
      \centering
      \includegraphics{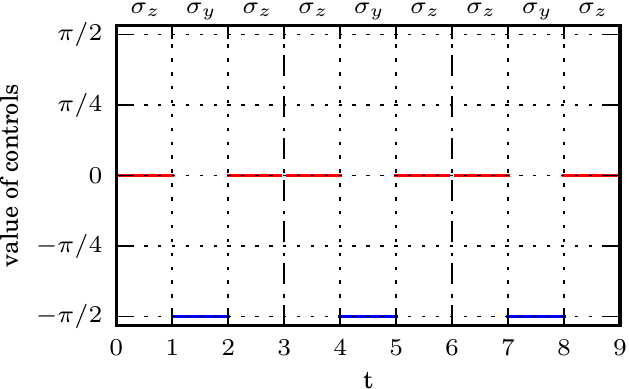}
      \label{pd_:b}
    }%
    \\
    \subfigure[Mean values of $\sigma_x,\sigma_y$ and $\sigma_z$.
     ]{%
      \centering
      \includegraphics{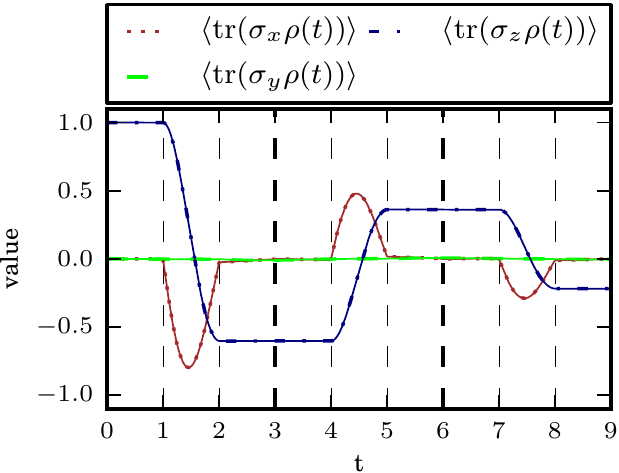}
      \label{pd_:c}
    }%
    \subfigure[Mean values of $\sigma_x,\sigma_y$ and $\sigma_z$.
         ]{%
      \centering
      \includegraphics{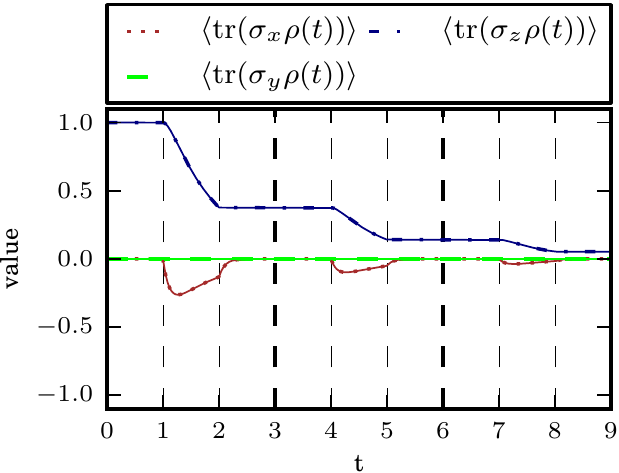}
      \label{pd_:d}
    }%
    \\
    \subfigure[Time evolution of a quantum coin. 
         ]{%
    	\centering
      \includegraphics{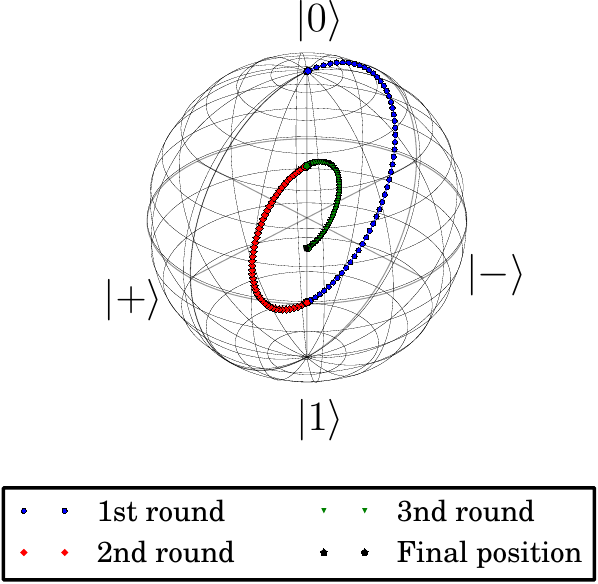}
      \label{pd_:e}
    }%
    \subfigure[Time evolution of a quantum coin. 
         ]{%
      \centering
      \includegraphics{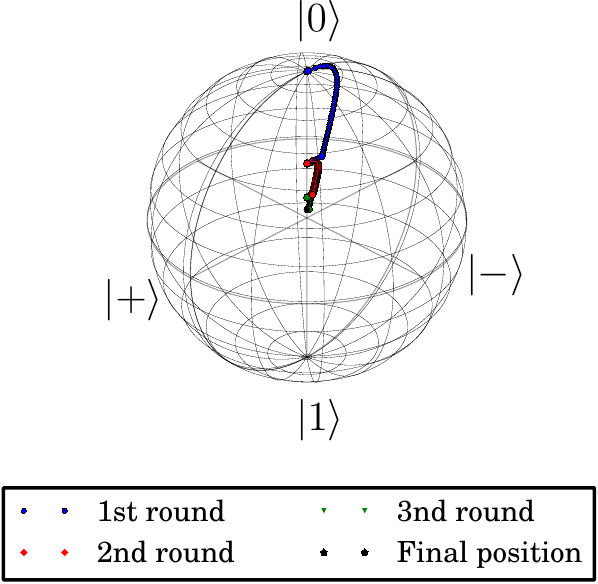}
      \label{pd_:f}
    }%

  \caption{Example of the time evolution of a quantum system with the phase
damping decoherence for fixed $\gamma=0.5$ (left side), $\gamma=5$ (right side) and a sequence of control parameters $\alpha$.} %
\label{fig:evolution_pd}
\end{figure}

Now we consider the impact of the phase damping decoherence on the outcome
of the game. In this case, the Lindblad operator is given by $\sigma_z$.
Hence, the Lindblad equation has the following form
\begin{equation}
\frac{\dd\rho}{\dd t}=
-\ii[H,\rho] + \gamma(\sigma_z\rho\sigma_z
- \frac{1}{2}\sigma_z\sigma_z\rho
-\frac{1}{2}\rho\sigma_z\sigma_z)=-\ii[H,\rho]
+ \gamma(\sigma_z\rho\sigma_z - \rho).
\end{equation}
Next, we make a change of variables 
$\hat{\rho}(t)=e^{\mathrm{i}Ht}\rho(t)e^{-\mathrm{i}Ht}$, which is helpful to
solve the equation. We obtain
\begin{equation}
\frac{\dd\hat{\rho}}{\dd t}=\gamma (\sigma_z\hat{\rho(t)}\sigma_z - \hat{\rho(t)}).
\end{equation}
It follows that the solution of the above equation is given by
\begin{equation}
\begin{split}
\hat{\rho}(t)=&\ketbra{0}{0}\rho(0)\ketbra{0}{0} +
\ketbra{1}{1}\rho(0)\ketbra{1}{1} + \\ +&  \mathrm{e}^{-2\gamma t}
(\ketbra{0}{0}\rho(0)\ketbra{1}{1}+\ketbra{1}{1}\rho(0)\ketbra{0}{0}).
\end{split}
\end{equation}
Coming back to the original variables we get the expression
\begin{equation}
\begin{split}
\rho(t)=&\ketbra{0}{0}\rho(0)\ketbra{0}{0} + \ketbra{1}{1}\rho(0)\ketbra{1}{1}
+ \\ +&
\mathrm{e}^{-2\gamma t}\mathrm{e}^{-\mathrm{i} H t}(\ketbra{0}{0}\rho(0)\ketbra{1}{1}+\ketbra{1}{1}\rho(0)\ketbra{0}{0}) \mathrm{e}^{\mathrm{i} H t}.
\end{split}
\end{equation}
Consider the following limit
\begin{equation}
\lim_{\gamma\to\infty} \rho(t)= \ketbra{0}{0}\rho(0)\ketbra{0}{0} + \ketbra{1}{1}\rho(0)\ketbra{1}{1}.
\end{equation}
The above result is a diagonal matrix dependent on the initial state. For high
values of $\gamma$, the initial state $\rho(0)$ has a significant impact on the
game. If $\rho(0)=\ketbra{0}{0}$ then $\lim_{\gamma\to\infty}
\rho(t)=\ketbra{0}{0}$. This kind of decoherence is conducive to Bob.
Similarly, if $\rho(0) = \ketbra{1}{1}$, then Alice wins. The evolution of a
quantum  system with the phase damping decoherence and fixed Hamiltonian is
shown in Figure~\ref{fig:evolution_pd}.

\section{Optimal strategy for the players}\label{sec:strategies}
Due to the noisy evolution of the underlying qubit, the strategy given by
Table~\ref{tab:control-parameter} is no longer a Nash equilibrium. We study the
possibility of optimizing one player's strategy, while the other one uses the
Pauli strategy. It turns out that this optimization is not always  possible. If
the rate of decoherence is high enough, then the players' strategies have
little impact on the game outcome. In the low noise scenario, it is  possible
to optimize the strategy of both players.

In each round, one player performs a series of unitary operations, which are
chosen randomly from a uniform distribution. Therefore, the strategy of a
player can be seen as a random unitary channel. In this section
$\Phi_{A_1},\Phi_{A_2}$ denote mixed unitary channels used by Alice who
implements the Pauli strategy. Similarly, $\Phi_B$ denotes channels used by
Bob.

\subsection{Optimization method}
In order to find optimal strategies for the players we assume the Hamiltonian
in~\eqref{equ:master-equation} to have the form
\begin{equation}
	H = H(\varepsilon(t)),
\end{equation}
where $\varepsilon(t)$ are the control pulses. As the optimization target, we
introduce the cost functional
\begin{equation}
	J(\varepsilon)=\tr\{ F_0(\rho(T)) \}\label{equ:cost},
\end{equation}
where $F_0(\rho(T))$ is a functional that is bounded from below and
differentiable with respect to $\rho(T)$. A sequence of control pulses that
minimizes the functional~\eqref{equ:cost} is said to be \emph{optimal}. In our
case we assume that
\begin{equation}
	\tr\{ F_0(\rho(T)) \} = \frac12 || \rho(T) - \rho_{\rm T} ||_{\rm F}^2,
\end{equation}
where $\rho_{\rm T}$ is the target density matrix of the system.

In order to solve this optimization problem, we need to find an analytical
formula for the derivative of the cost functional~\eqref{equ:cost} with respect
to control pulses $\varepsilon(t)$. Using the Pontryagin
principle~\cite{pontryagin1962mathematical}, it is possible to show that we
need to  solve the following equations to obtain the analytical formula for the
derivative~\cite{jirari2005optimal}
\begin{align}
	\frac{\dd\rho(t)}{\dd t}&=-\ii[H(\varepsilon(t))
	,\rho(t)] - \ii L_{\rm D} [\rho(t)],\; t\in[0,
	T]\label{equ:forward},\\
	\frac{\dd \lambda(t)}{\dd t} &= -\ii [H(\varepsilon(t))
	,\lambda(t)] - \ii L_{\rm D}^\dagger [\lambda(t)],\; t\in[0,
	T]\label{equ:backward},\\
	L_{\rm D}[A] &= \ii\sum_j \gamma_j(L_j A L_j^\dagger -
	\frac{1}{2}\{L_j^\dagger L_j,A\}),\\
	\rho(0) &= \rho_{\rm s}, \\
	\lambda(T) &= F'_0(\rho(T)),
\end{align}
where $\rho_{\rm s}$ denotes the initial density matrix, $\lambda(t)$ is called
the adjoint state and
\begin{equation}
	F'_0(\rho(T)) = \rho(T) - \rho_{\rm T}.
\end{equation}

In order to optimize the control pulses using a gradient method, we convert the
 problem from an infinite dimensional (continuous time) to a finite dimensional
 (discrete time) one. For this purpose, we discretize the time interval $[0,
T]$ into $M$ equal sized subintervals $\Delta t_k$. Thus, the problem becomes
that of finding
$\varepsilon=[\varepsilon_1,\ldots,\varepsilon_M]^{\rm T}$  such that
\begin{equation}
	J(\varepsilon) = \inf_{\zeta \in \mathbb{R}^M}J(\zeta).
\end{equation}
The gradient of the cost functional is
\begin{equation}
	G = \left[ \frac{\partial J}{\partial \varepsilon_1}, \ldots,
	\frac{\partial J}{\partial \varepsilon_M} \right]^{\rm
	T}\label{equ:gradient}.
\end{equation}
It can be shown~\cite{jirari2005optimal} that elements of
vector~\eqref{equ:gradient} are given by
\begin{equation}
	\frac{\partial J}{\partial \varepsilon_k} = \tr \left\{ -\ii \lambda_k
	\left[ \frac{\partial H(\varepsilon_k)}{\partial \varepsilon_k},
	\rho_k \right] \right\}\Delta t_k,
\end{equation}
where $\rho_k$ and $\lambda_k$ are solutions of the Lindblad equation and
the adjoint system corresponding to time subinterval $\Delta t_k$ respectively.
To minimize the gradient given in Eq.~\eqref{equ:gradient} we use the BFGS
algorithm~\cite{press1992numerical}.

\subsection{Optimization setup}
Our goal is to find control strategies for players, which maximize their
respective chances of winning the game. We study three noise channels: the
amplitude damping, the phase damping and the amplitude raising channel. They
are given by the Lindblad operators $\sigma_-$, $\sigma_z$ and $\sigma_+ =
\sigma_-^\dagger$, respectively. In all cases, we assume that one of the
players uses the Pauli strategy, while for the other player we try to optimize
a control strategy that maximizes that player's probability of winning.
However, in our setup it is convenient to use the value of the observable
$\sigma_z$ rather than probabilities. Value 0 means that each player has a
probability of $\frac12$ of winning the game. Values closer to 1 mean higher
probability of winning for Bob, while values closer to -1 mean higher
probability of winning for Alice.

\subsection{Optimization results}
\subsubsection{Phase damping}
\begin{figure}[ht!]
	\centering
	\includegraphics{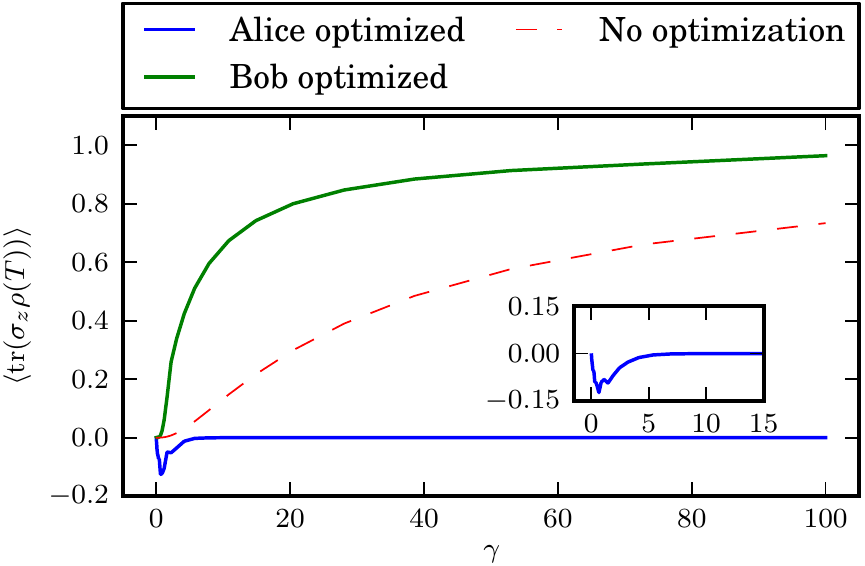}
\caption{Mean value of the pay-off for the phase damping channel with and
without optimization of the player's strategies. The inset shows the region
where Alice is able to increase her probability of winning to exceed
$\frac12$.}\label{fig:phase_damp}
\end{figure}

\begin{figure}[ht!]
   \centering
    \subfigure[Optimal controls for Alice.
     ]{%
      \centering
      \includegraphics{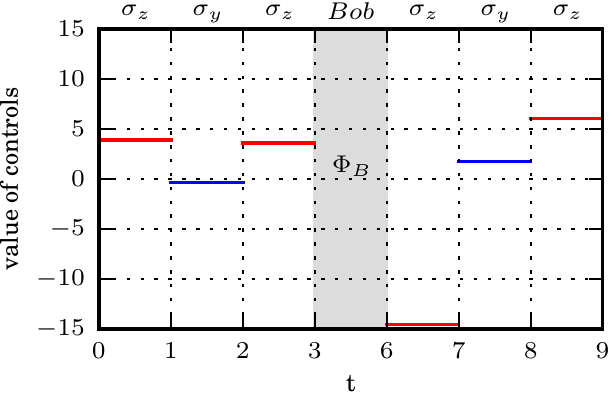}
      \label{pd:a}
    }%
    \subfigure[Optimal controls for Bob.
     ]{%
      \centering
      \includegraphics{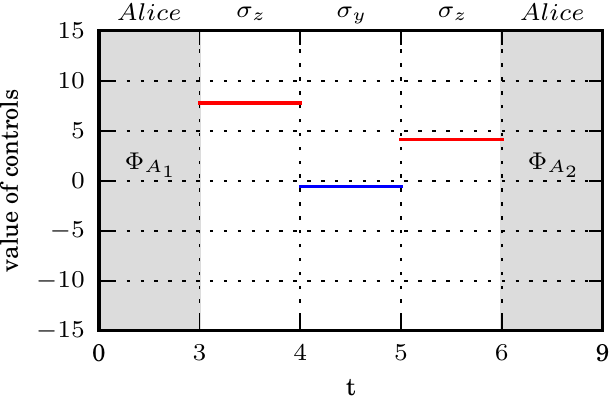}
      \label{pd:b}
    }%
    \\
    \subfigure[Mean values of $\sigma_x,\sigma_y$ and $\sigma_z$.
     ]{%
      \centering
      \includegraphics{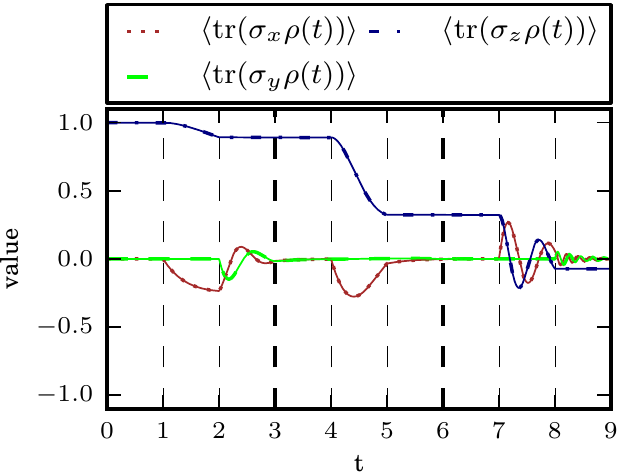}
      \label{pd:c}
    }%
    \subfigure[Mean values of $\sigma_x,\sigma_y$ and $\sigma_z$. 
     ]{%
      \centering
      \includegraphics{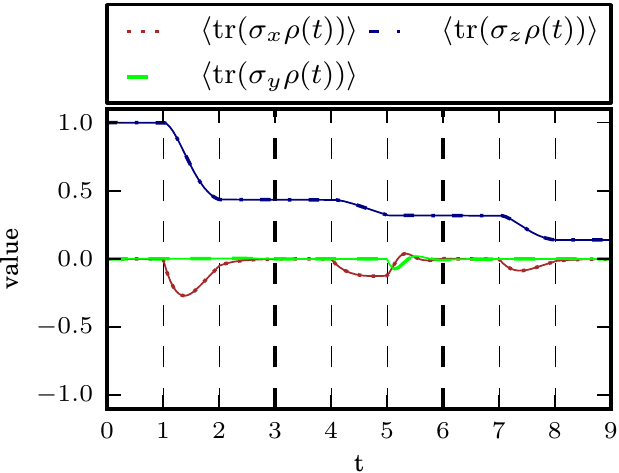}
      \label{pd:d}
    }%
    \\
    \subfigure[Time evolution of a quantum coin.
         ]{%
    	\centering
      \includegraphics{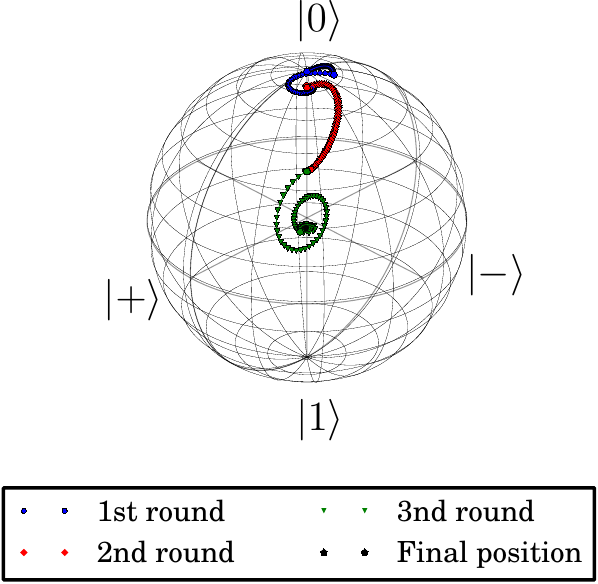}
      \label{pd:e}
    }%
    \subfigure[Time evolution of a quantum coin.
     ]{%
      \centering
      \includegraphics{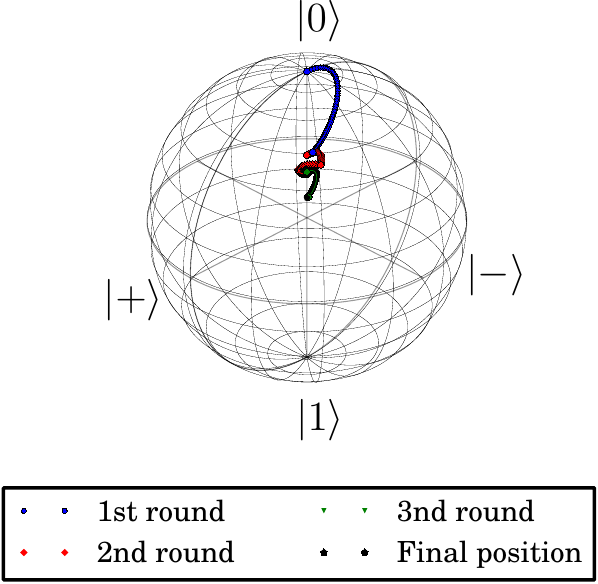}
      \label{pd:f}
    }%

  \caption{Game results for the phase damping channel. Optimal Alice's
  strategy when $\gamma = 1.172$ (left side), and optimal Bob's strategy when
  $\gamma = 1.610$ (right side).} %
  \label{fig:opt_pd}
\end{figure}

The results for the phase damping channel are shown in
Figure~\ref{fig:phase_damp}. As it can be seen, in this case both players are
able to  optimize their strategies, and so Alice can optimize her strategy for
low values of $\gamma$ to obtain the probability of winning grater than
$\frac12$.  The region where this occurs is shown in the inset. For high noise
values she is able to achieve the probability of winning equal to $\frac12$. On
the other hand, optimization of Bob's strategy shows that he is able to achieve
high probabilities of winning for relatively  low values of $\gamma$.
Figure~\ref{fig:opt_pd} presents optimal game strategies for both  players. For
Alice we chose $\gamma=1.172$ which corresponds to her maximal  probability of
winning the game. In the case of Bob's strategies we  arbitrarily choose the
value $\gamma=1.610$.

\subsubsection{Amplitude damping}
\begin{figure}[!h]
	\centering
	\includegraphics{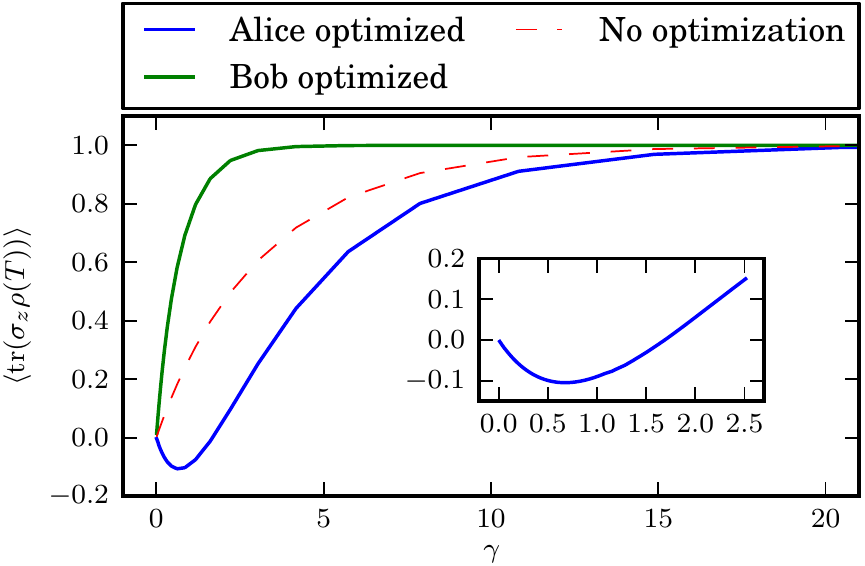}
\caption{Mean value of the pay-off for the amplitude damping channel with and
without optimization of the player's strategies. The inset shows the region
where Alice is able to increase her probability of winning to exceed
$\frac12$.}\label{fig:amp_damp}

\end{figure}

\begin{figure}[h!]
   \centering
     \subfigure[Optimal controls for Alice.
     ]{%
			 \centering
       \includegraphics{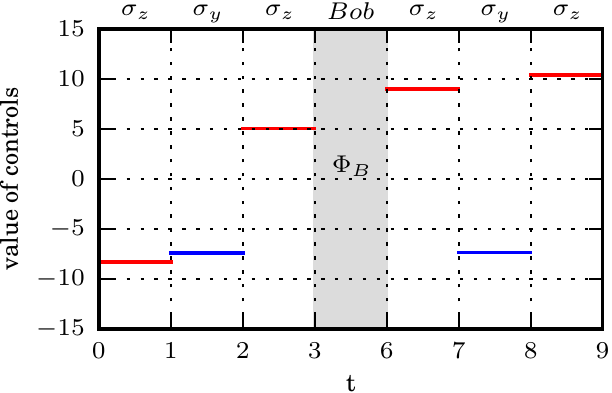}
       \label{ad:a}
     }%
     \subfigure[Optimal controls for Bob. 
     ]{%
     	 \centering
       \includegraphics{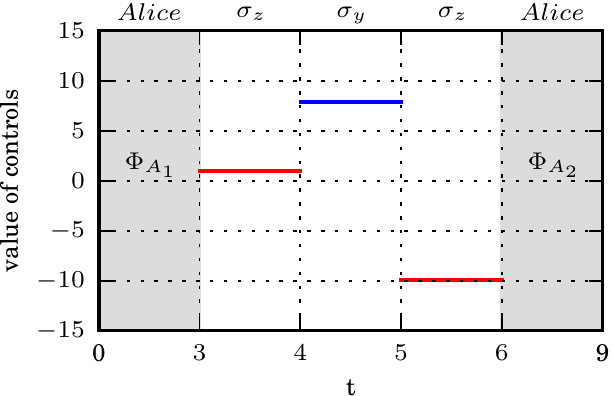}
       \label{ad:b}
     }%
     \\
    \subfigure[Mean values of $\sigma_x,\sigma_y$ and $\sigma_z$.
     ]{%
			\centering
      \includegraphics{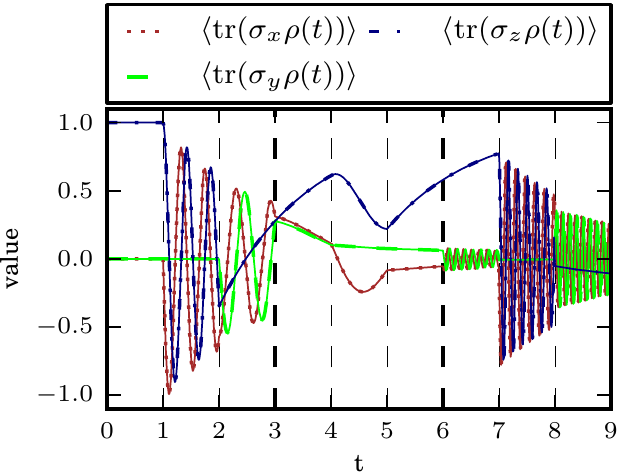}
      \label{ad:c}
    }%
    \subfigure[Mean values of $\sigma_x,\sigma_y$ and $\sigma_z$. 
     ]{%
			\centering
      \includegraphics{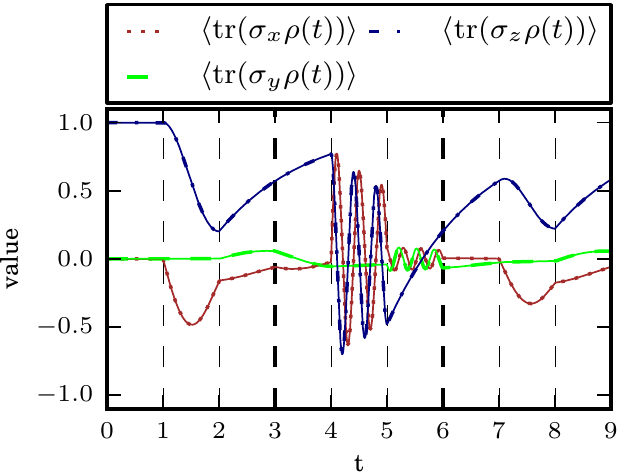}
      \label{ad:d}
    }%
    \\
    \subfigure[Time evolution of a quantum coin.
         ]{%
  		\centering
      \includegraphics{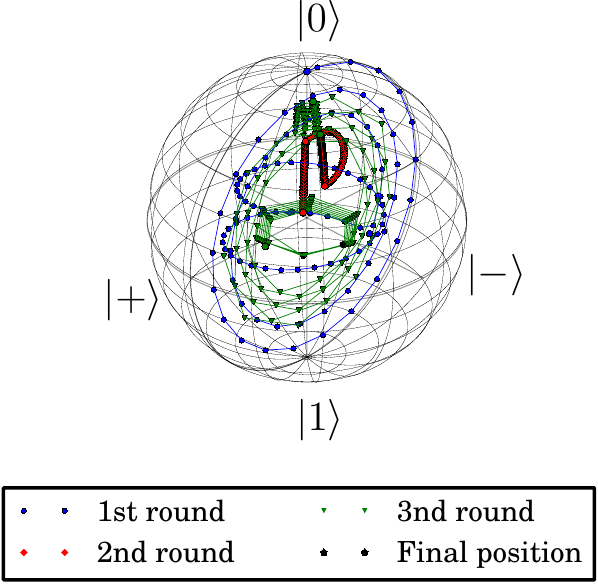}
      \label{ad:e}
    }%
    \subfigure[Time evolution of a quantum coin.
	    	 ]{%
			\centering
      \includegraphics{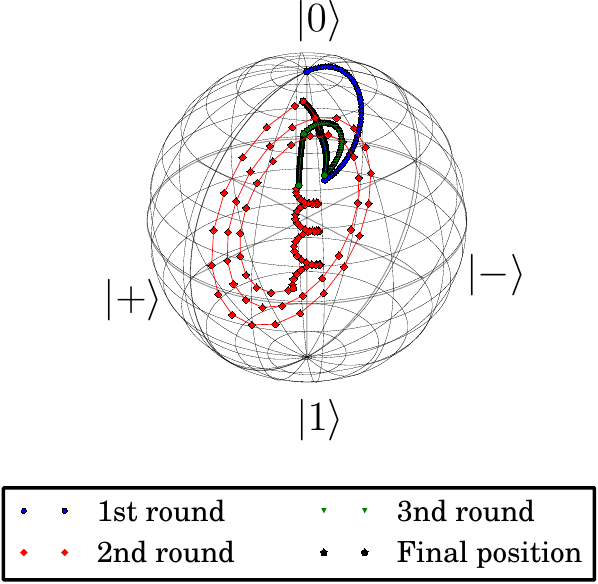}
      \label{ad:f}
    }%

  \caption{Game results obtained for the amplitude damping channel with
$\gamma$ equal to $0.621$. Optimal Alice's strategy (left side), and optimal
Bob's strategy (right side).} %
\label{fig:opt_ad} 
\end{figure}

Next, we present the results obtained for the amplitude damping channel. They
are shown in Figure~\ref{fig:amp_damp}. Unfortunately for Alice, for high
values of $\gamma$ Bob always wins. This is due to the fact that in this case
the state quickly decays to state $\ketbra{0}{0}$. Additionally, Bob is also
able to optimize his strategies. He is able to achieve probability of winning
equal to 1 for relatively low values of $\gamma$. For low values of $\gamma$
the interaction allows Alice to achieve higher than $\frac12$ probability of
winning. The region where this happens is magnified in the inset.
Interestingly, for very low values of $\gamma$ Alice can increase her
probability of winning. This is due to the fact, that low noise values are
sufficient to distort Bob's attempts to perform the Pauli strategy. On the
other hand, they are not high enough to drive the system towards state
$\ketbra{0}{0}$. Optimal game results for both players are shown in
Figure~\ref{fig:opt_ad}. For both players we chose $\gamma=0.621$ which
corresponds to Alice's maximal probability of winning the game.

\subsubsection{Amplitude raising}
\begin{figure}[!ht]
	\centering
	\includegraphics{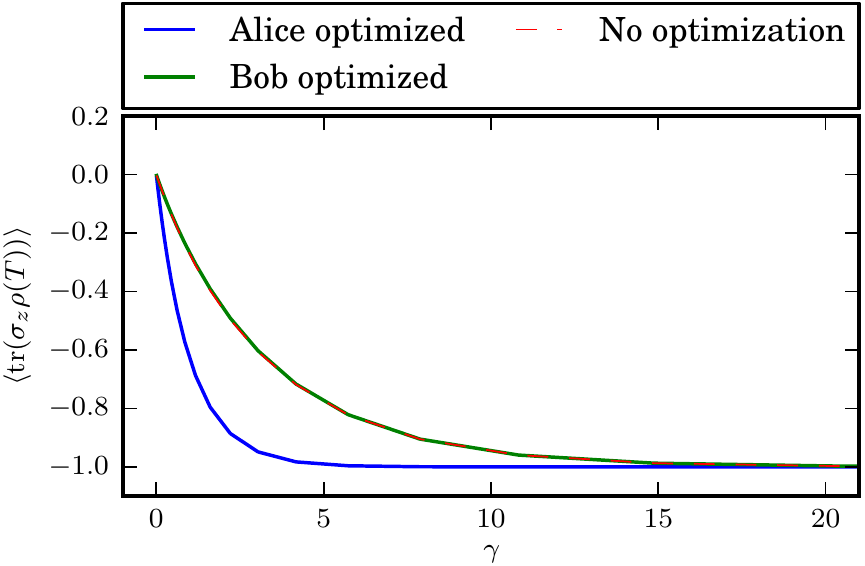}
\caption{Mean value of the pay-off for the amplitude raising channel with and
without optimization of the player's strategies.}\label{fig:amp_raise}
\end{figure}

Finally, we present optimization results for the amplitude raising channel.
The optimization results, shown in  Figure~\ref{fig:amp_raise}, indicate that
Alice can achieve probability of  winning equal to 1 for lower values of
$\gamma$ compared to the unoptimized  case. In this case Bob cannot do any
better than in the unoptimized case  due to a limited number of available
control pulses.

\section{Conclusions}\label{sec:conclusions}
We studied the quantum version of the coin flip game under decoherence. To
model  the interaction with external environment we used the Markovian
approximation  in the form of the Lindblad equation. Because of the fact that
Pauli strategy is a known Nash equilibrium of the game, therefore it was
natural to investigate this strategy in the presence noise. Our  results show
that in the presence of noise the Pauli strategy is no longer a  Nash
equilibrium. One of the players, Bob in our  case, is always favoured by
amplitude and phase damping noise.

Our next step was to check if the players were able to do better than the Pauli
strategy. For this, we used the BFGS gradient method to optimize the players'
strategies. Our results show that Alice, as well as Bob, are able to increase
their respective winning probabilities. Alice can achieve this for all  three
studied cases, while Bob can only do this for the phase damping and  amplitude
damping channels.
\section*{Acknowledgements}
The work was supported by the Polish Ministry of Science and
Higher Education grants:
P.~Gawron under the project number IP2011 014071.
D.~Kurzyk under the project number N N514 513340.
\L{}.~Pawela under the project number N N516 481840.

\bibliographystyle{plain}
\bibliography{qubit_flip_lindblad}
\end{document}